\documentstyle[epsfig]{aa}

\newcommand{\be}{\begin{equation}}
\newcommand{\ee}{\end{equation}}

\begin{document}

\title{GRB afterglow light curves from uniform and non-uniform jets}
\author{D.M. Wei\inst{1,2} and Z.P. Jin\inst{1,2}}
\institute{Purple Mountain Observatory, Chinese Academy of Sciences,
Nanjing, China
\and
   National Astronomical Observatories, Chinese Academy of Sciences,
   China}
\date{Received date/ Accepted date}

\abstract{It is widely believed that gamma-ray bursts are produced by a
jet-like outflows directed towards the observer, and the jet opening
angle($\theta_j$) is often inferred from the time at which there is a
break in the afterglow light curves. Here we calculate the GRB
afterglow light curves from a relativistic jet as seen by observers at
a wide range of viewing angles ($\theta_v$) from the jet axis, and the
jet is uniform or non-uniform(the energy per unit solid angle decreases
smoothly away from the axis $\epsilon (\theta)\propto
(\theta/\theta_c)^{-k}$). We find that, for uniform jet($k=0$), the
afterglow light curves for different viewing angles are somewhat
different: in general, there are two breaks in the light curve, the
first one corresponds to the time at which $\gamma\sim
(\theta_j-\theta_v)^{-1}$, and the second one corresponds to the time
when $\gamma\sim (\theta_j+\theta_v)^{-1}$. However, for non-uniform
jet, the things become more complicated. For the case $\theta_v=0$, we
can obtain the analytical results, for $k<8/(p+4)$(where $p$ is the
spectral index of electron energy distribution) there should be two
breaks in the light curve correspond to $\gamma\sim\theta_c^{-1}$ and
$\gamma\sim\theta_j^{-1}$ respectively, while for $k>8/(p+4)$ there
should be only one break corresponds to $\gamma\sim\theta_c^{-1}$, and
this provides a possible explanation for some rapidly fading afterglows
whose light curves have no breaks since the time at which
$\gamma\sim\theta_c^{-1}$ is much earlier than our first observation
time. For the case $\theta_v\neq 0$, our numerical results show that,
the afterglow light curves are strongly affected by the values of
$\theta_v$, $\theta_c$ and $k$. If $\theta_v$ is close to $\theta_c$
and $k$ is small, then the light curve is similar to the case of $k=0$,
except the flux is somewhat lower. However, if the values of
$\theta_v/\theta_c$ and $k$ are larger, there will be a prominent
flattening in the afterglow light curve, which is quite different from
the uniform jet, and after the flattening a very sharp break will be
occurred at the time $\gamma\sim (\theta_v + \theta_c)^{-1}$.
} \maketitle

\section{Introduction}

Gamma-ray bursts (GRBs) are known as an explosive phenomenon occurring
at cosmological distances, emitting large amount of energy mostly in
the gamma-ray range (see, e.g.  Piran 1999; Cheng \& Lu 2001 for a
review). Observations show that some of GRBs are emitting an extremely
large energy with $E_\gamma\gg 10^{52}$ ergs if emission is isotropic.
For example, GRB990123, the most energetic GRB event detected so far,
has an isotropic gamma-ray energy of $\sim 3.4\times 10^{54}$ ergs,
which corresponds to the rest-mass energy of $\sim 1.9M_{\odot}$
(Kulkarni et al. 1999). Such a crisis of extreme large energy forced
some people to think that the GRB emission must be highly collimated in
order to reduce the total energy.

The second reason for GRB emission being jet-like comes from the fact
that there are sharp breaks in the light curves of some GRBs'
afterglows, such as GRB990123 (Kulkarni et al. 1999; Castro-Tirado et
al. 1999) and GRB990510 (Harrison et al. 1999; Stanek et al. 1999),
etc.. These observed breaks have generally been interpreted as evidence
for collimation of the GRB ejecta, since Rhoads (1999) and Sari et al.
(1999) have pointed out that the lateral expansion of the relativistic
jet can produce a sharp break in the afterglow light curve.

However, in the current afterglow jet models, it is generally assumed
that the jet is uniform, and the line-of-sight is just along the jet
axis. It is obvious that these assumptions are usually not true, since
some GRB models predict that the jet may be non-uniform, within which
the energy per unit solid angle decreases away from the jet axis, such
as $\epsilon\propto\theta^{-k}$ (e.g. MacFadyen et al. 2001), and the
probability that the line-of-sight just crosses the jet axis is near
zero. So it is very important to investigate the situation that the jet
is non-uniform and is seen by observers at a wide range of viewing
angles from the jet axis. Some authors have already considered the
situation of anisotropic jets (Meszaros, Rees \& Wijers 1998; Salmonson
2001; Dai \& Gou 2001; Zhang \& Meszaros 2002; Rossi, Lazzati \& Rees
2002). In this paper, we give a detailed calculation of the emission
from anisotropic jets, including the effect of equal-arrival-time
surface. In the next section we discuss the dynamical evolution of the
jet, in section 3 we calculate the emission from uniform jet for
different viewing angle, in section 4 we calculate the emission
features from non-uniform jet, and finally we present some discussions
and conclusions.

\section{Dynamical evolution of the jet}

Now we consider an adiabatic relativistic jet expanding in the
surrounding medium. For energy conservation, the evolution equation is
\be \gamma^{2}\Delta Nm_pc^2=\epsilon\Delta\Omega \ee where $\gamma$ is
the bulk Lorentz factor, $\Delta N=\frac{1}{3}nr^{3}\Delta\Omega$ is
the particle numbers swept by the jet within a solid angle
$\Delta\Omega$, $n$ is the surrounding medium density, and $\epsilon$
is the energy per unit solid angle.

It is well known that for relativistic blast waves, the received
photons at time $T$ are not emitted at the same time. A photon that is
located at radius $r$ and is emitted with an angle $\Theta$ from the
line-of-sight will reach the observer at time $T=r(1-\beta\mu)/(\beta
c)$, where $\mu=\cos\Theta$. Then we can obtain the jet evolution \be
\gamma=(\frac{3\epsilon}{n})^{1/2}(m_{p}c^{2})^{-1/2}(\frac{\beta cT}
{1-\beta\mu})^{-3/2} \ee We see that the jet Lorentz factor evolution
is dependent on the angle $\Theta$, for different values of $\Theta$
the relation between $\gamma$ and $T$ are different. Here we neglect
the sideways expansion of the jet since this process is very unclear.
For relativistic jet, \be \gamma (1-\beta\mu)^{-3/2}\simeq 3.4\times
10^{5}(\frac{\epsilon_{54}}{n_1})^{1/2}T_{day}^{-3/2} \ee where
$\epsilon_{54}$ is the energy in units of $10^{54}$ ergs, $n_1$ is the
surrounding density in units of 1 atom $cm^{-3}$, and $T_{day}$ is the
observed time in units of 1 day. In the case of $\gamma\gg 1$ and
$\Theta\ll 1$, we have \be \gamma(1+\gamma^2\Theta^2)^{-3/8}=
18.6(\frac{\epsilon_{54}}{n_1})^{1/8}T_{day}^{-3/8} \ee This equation
can be solved numerically, and by fitting the numerical results, we
obtain the analytical solution \be \gamma=\left\{\begin{array}{ll}
10^{1.27}(\frac{\epsilon_{54}}{n_1})^{1/8}T_{day}^{-3/8}, & {\rm for}
\;\;\;\gamma\Theta\leq 1 \\
10^{5.08}(\frac{\epsilon_{54}}{n_1})^{1/2}T_{day}^{-3/2}\Theta^{3}, &
{\rm for} \;\;\;\gamma\Theta > 1
\end{array} \right . \ee
It is obvious that the usual solution $\gamma\propto T^{-3/8}$ is valid
only when $\Theta \leq \gamma^{-1}$, while when $\Theta > \gamma^{-1}$
the relation becomes $\gamma\propto T^{-3/2}$.

\section{emission from uniform jet}

Emission features from uniform jet has been discussed by many authors,
and it is widely believed that there is a sharp break in the GRB
afterglow light curve corresponds to the time $\gamma\sim
\theta_j^{-1}$, where $\theta_j$ is the jet half-opening angle.
However, this is true only when the observer line-of-sight just crosses
the jet axis, and in fact this probability is very small. So here we
calculate the jet emission for various viewing angles.

We follow our previous paper (Wei \& Lu 2000) to calculate the emission
flux from the jet. We assume that the line-of-sight is along $z$-axis,
the symmetry axis of the jet is in the $y-z$ plane, $\theta_v $ is the
angle between the line-of-sight and the symmetry axis, and the
radiation is isotropic in the comoving frame of the ejecta. In order to
see more clearly, let us establish an auxiliary coordinate system ($x',
y', z'$) with the $z'$-axis along the symmetry axis of the cone and the
$x'$ parallel the $x$-axis. Then the position within the cone is
specified by its angular spherical coordinates $\theta $ and $\phi $
($0\leq \theta \leq \theta _{j}$, $0\leq \phi \leq 2\pi $). It can be
shown that the angle $\Theta $ between a direction ($\theta ,\,\phi $)
within the cone, and the line-of-sight satisfies $cos\Theta
=cos\theta_{v} cos\theta -sin\theta_{v} sin\theta sin\phi $. Then the
observed flux is \be F_{\nu}=\int_{0}^{2\pi }d\phi \int_{0}^{\theta
_{j}}sin\theta d\theta D ^{3} I'(\nu D^{-1})\frac{r^{2}}{d^{2}} \ee
where $D=[\gamma (1-\beta cos\Theta )]^{-1}$ is the Doppler factor,
$\beta =(1-\gamma ^{-2})^{1/2}$, $\nu =D\nu '$, $I'(\nu ')$ is the
specific intensity of synchrotron radiation at $\nu '$, and $d$ is the
distance of the burst source. Here the quantities with primes are
measured in the comoving frame.

It is generally believed that the electrons have been accelerated by
the shock to a power law distribution $n_{e}(\gamma_e )\propto \gamma_e
^{-p}$ for $\gamma _{min}\leq \gamma_e \leq \gamma _{max}$, and
consider the synchrotron radiation of these electrons, we can obtain
the observed flux \be F_{\nu} \propto T^{3}\nu ^{-\frac{p-1}{2}}
\int_0^{2\pi }d\phi \int_{0}^{\theta _{j}}sin\theta d\theta
f(\gamma)(1-\beta\mu)^{-\frac{p+11}{2}} \ee where $T$ is the time
measured in the observer frame, $f(\gamma )=\gamma ^{-(p+7)/2}(\gamma
-1) ^{(p+1)/4}y(\gamma )^{(p+5)/4}\gamma _{min}^{p-1}\beta ^{3}$,
$y(\gamma )=\frac{\hat {\gamma }\gamma +1}{\hat {\gamma }-1}$, $\hat
{\gamma }$ is just the ratio of specific heats, $\gamma _{min}=\xi _{e}
(\gamma -1)\frac{m_{p}}{m_{e}}\frac{p-2}{p-1}$, $m_{p}(m_{e})$ is the
mass of proton (electron). In the relativistic case, we have \be
F_{\nu} \propto T^{3}\nu ^{-\frac{p-1}{2}} \int_0^{2\pi }d\phi
\int_{0}^{\theta _{j}}sin\theta d\theta \gamma^{p-3}
(1-\beta\mu)^{-\frac{p+11}{2}} \ee Combining the jet evolution results
(eq. (2)), finally we obtain the observed flux \be F_{\nu} \propto
T^{-\frac{3}{2}p+\frac{15}{2}}\nu ^{-\frac{p-1}{2}} \int_0^{2\pi }d\phi
\int_{0}^{\theta _{j}}sin\theta d\theta \epsilon^{\frac{p-3}{2}}
(1-\beta\mu)^{p-10} \ee This is the basic equation for our calculation.
Here we take $\theta_j=0.1$, $p=2.5$ and $n_1=1$ throughout this paper.

For uniform jet, the energy per unit solid angle $\epsilon$ is
independent of $\theta$, $\phi$, we take $\epsilon_{54}=1$. Then using
equations (2) and (9), we can calculate the jet emission for different
viewing angles. Fig.1 gives our numerical results. From Fig.1 we see
that the afterglow light curves for different viewing angles are
somewhat different: in general, there are two breaks in the light
curve, the first one corresponds to the time at which $\gamma\sim
(\theta_j-\theta_v)^{-1}$, and the second one corresponds to the time
when $\gamma\sim (\theta_j+\theta_v)^{-1}$. This is quite different
from the previous results, which think there is one break occurred at
$\gamma\sim \theta_j^{-1}$.

\begin{figure}
\epsfig{file=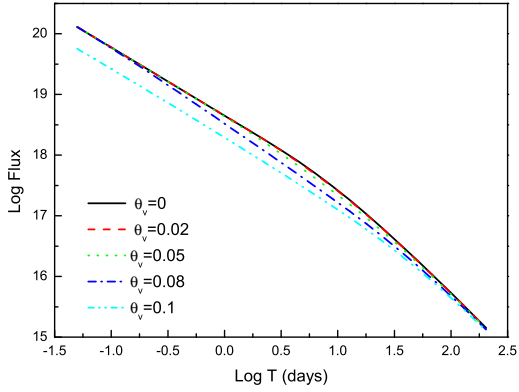, width=8cm} \caption{The afterglow light curves
from uniform jet for various viewing angles.}
\end{figure}

\section{Emission from non-uniform jet}

Uniform jet is only a special case, in general we expect the jet is
non-uniform, for instance, the collapsar model predicted that the jet
is non-uniform, within which the energy per unit solid angle decreases
away from the jet axis (e.g. MacFadyen et al. 2001). Here we suppose
the energy per unit solid angle has the form \be
\epsilon=\left\{\begin{array}{ll} \epsilon_{0} & \;\;{\rm for}
\;\;\;\;\;0\leq \theta \leq \theta_{c} \\
\epsilon_{0}(\frac{\theta}{\theta_c})^{-k} & \;\;{\rm for}
\;\;\;\;\;\theta_c \leq \theta \leq \theta_j
\end{array} \right . \ee
where $\theta_c$ is introduced to avoid a divergence at $\theta=0$. For
arbitrary viewing angle $\theta_v$, the situation is very complicated,
so first we consider the case $\theta_v=0$, since this can give an
analytical analysis.

In the case of $\theta_v=0$, we have $\mu=\cos\theta$. It should be
noted that there is an interesting phenomenon, the values of
$\gamma\theta$ increase with the angle $\theta$, i.e. there is a
characteristic angle $\theta_{\star}$, at which the value
$\gamma_{\star}\theta_{\star}=1$, when $\theta<\theta_{\star}$ the
value $\gamma\theta<1$, and when $\theta> \theta_{\star}$ the value
$\gamma\theta>1$. It is obvious that the main contribution of emission
comes from the region $\theta\leq \theta_{\star}$, so $\theta_{\star}$
is an important quantity. It can be shown that when $T\leq T_1 \equiv
1.7(\frac{\epsilon_{0,54}}{n_1})^{1/3}(\frac{\theta_c}{0.02})^{8/3}$
hours, the value
$\theta_{\star}=10^{-1.27}(\frac{\epsilon_{0,54}}{n_1})^{-1/8}
T_{day}^{3/8}\leq \theta_c$, and when $T\leq T_2 \equiv 5.2\times
(0.2)^{k/3}(\frac{\epsilon_{0,54}}{n_1})^{1/3}
(\frac{\theta_c}{0.02})^{k/3}(\frac{\theta_j}{0.1})^{(8-k)/3}$ days,
the value $\theta_{\star}=10^{-1.27\times
8/(8-k)}(\frac{\epsilon_{0,54}}{n_1})
^{-1/(8-k)}\theta_c^{-k/(8-k)}T_{day}^{3/(8-k)}\leq \theta_j$.

Taking equation (5) and the approximate expression $1-\beta\mu \simeq
(2\gamma^2)^{-1}$ for $\theta<\theta_{\star}$, or $1-\beta\mu \simeq
\frac{1}{2}\theta^2$ for $\theta>\theta_{\star}$, we can get the
analytical results: (1) when $T\leq T_1$, the observed flux
$F_{\nu}\propto T^{-3(p-1)/4}$; (2) when $T_1\leq T \leq T_2$, the flux
$F_{\nu}\propto T^{-\frac{3(2p+k-2)}{8-k}}$ for $k<\frac{8}{p+4}$, or
$F_{\nu} \propto T^{-3p/4}$ for $k>\frac{8}{p+4}$; (3) when $T>T_2$,
the flux $F_{\nu}\propto T^{-3p/4}$. From this result we see that, for
smaller value of $k$, the transition of the afterglow light curve index
from $-\frac{3(p-1)}{4}$ to $-\frac{3p}{4}$ is gradual and smooth with
a timescale of about $T_2$, while for larger values of $k$, the
transition is rapid with a timescale of about $T_1$.

In order to verify the above results, we also make a numerical
calculation for the case $\theta_v=0$. Fig.2 gives our results. It is
obvious that the numerical results are consistent with the analytical
results, for larger value of $k$, the steepening of the light curve is
more rapidly. We suggest this may explain some afterglow light curves
which decay rapidly and have no breaks, since for larger value of $k$
the transition time ($\sim T_1$) is earlier than our first observation
time.

However, it should be noted that the appearance of the early break in
the light curve (corresponding to the time when $\gamma\sim
\theta_c^{-1}$) is due to the assumed energy distribution function
(equation (10)), and the sharpness of this break is primarily dependent
on the discontinuity in slope of $\epsilon$ for the idealized model of
equation (10) at $\theta=\theta_c$.  It is obvious that for a realistic
energy distribution, the transition of $\epsilon$ from roughly constant
for $\theta < \theta_c$ to $\epsilon\propto\theta^{-k}$ for $\theta >
\theta_c$ should be smooth. Therefore this break may be washed out by a
realistic energy distribution.

\begin{figure}
\epsfig{file=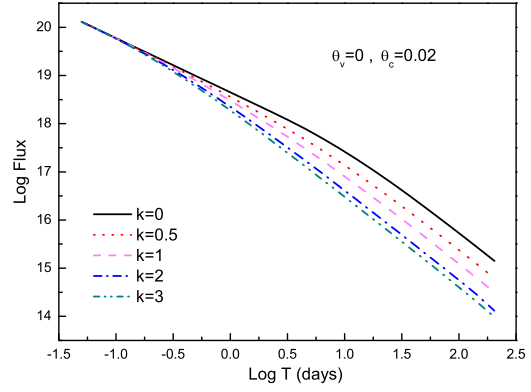, width=8cm} \caption{The afterglow light curves
from non-uniform jet for the case $\theta_v=0$.}
\end{figure}

For the general case $\theta_v\neq 0$, we calculate the afterglow light
curves numerically using equations (2), (9) and (10), the results are
given in Fig.3 - Fig.6. From these figures we see that the afterglow
light curves are dependent on the values of $\theta_v$, $\theta_c$ and
$k$. When $\theta_v$ is near $\theta_c$ and $k$ is small, then the
light curve is similar to the case of $\theta_v=0$, except the flux is
somewhat lower. However, if the viewing angle $\theta_v$ is larger than
$\theta_c$ and $k$ is not very small, then there will be a prominent
flattening in the afterglow light curve, which is quite different from
the case of uniform jet, and after the flattening a very sharp break
will be occurred around the time $\gamma\sim (\theta_v +
\theta_c)^{-1}$. If the viewing angle $\theta_v$ is larger than the jet
half-opening angle $\theta_j$, then the flux will first increase with
time until the Lorentz factor is about $\gamma \sim (\theta_v
-\theta_j)^{-1}$, thereafter the flux begin to decay with time.

\begin{figure}
\epsfig{file=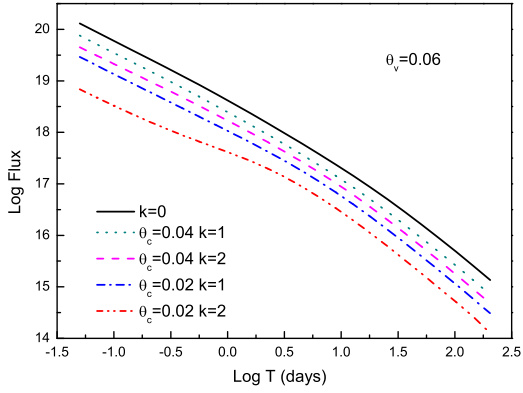, width=8cm} \caption{The afterglow light curves
from non-uniform jet for the case $\theta_v=0.06$.}
\end{figure}

\begin{figure}
\epsfig{file=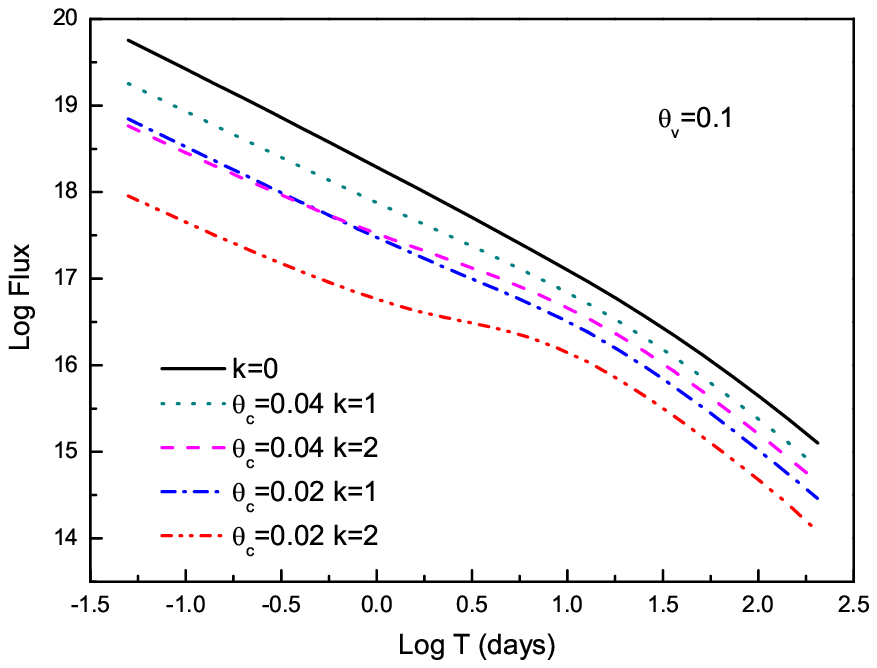, width=8cm} \caption{The afterglow light curves
from non-uniform jet for the case $\theta_v=0.1$.}
\end{figure}

\begin{figure}
\epsfig{file=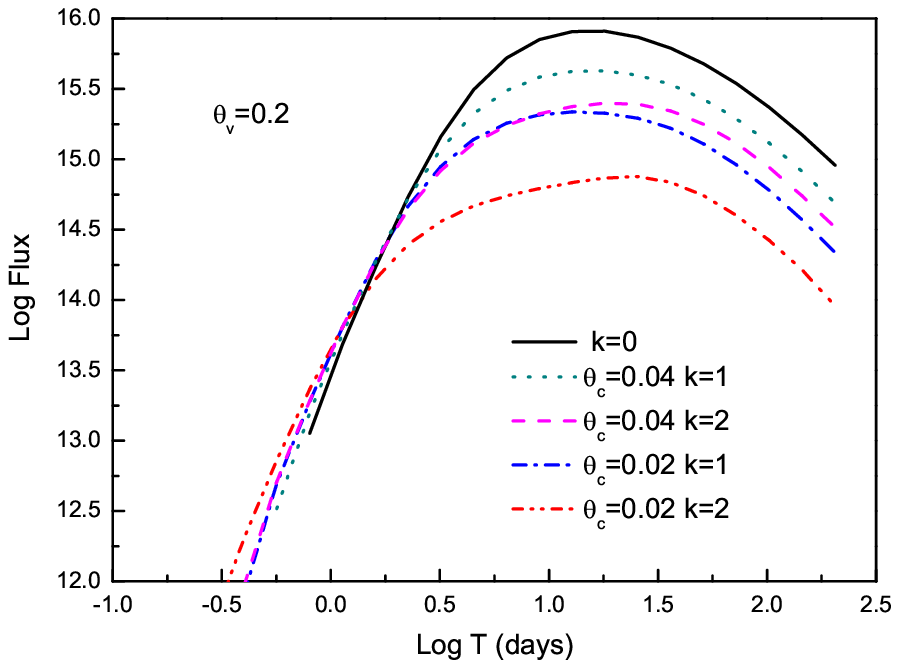, width=8cm} \caption{The afterglow light curves
from non-uniform jet for the case $\theta_v=0.2$.}
\end{figure}

\begin{figure}
\epsfig{file=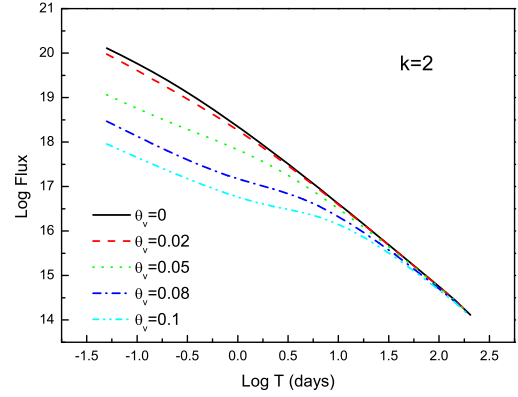, width=8cm} \caption{The afterglow light curves
from non-uniform jet ($k=2$) for different viewing angles.}
\end{figure}

\section{Discussion and conclusion}

In this paper we calculate the GRB afterglow light curves from
relativistic jets in more details, assuming that the jet may be uniform
or non-uniform, and the observer locate at arbitrary angle with respect
to the jet axis. We have shown that there are several distinct features
of our jet emission compared with previous jet model.

In previous analysis, it is generally assumed that the jet is uniform
and the line-of-sight is just along the jet axis, in this case the
afterglow light curves have a break at the time $\gamma\sim
\theta_j^{-1}$. However, if the viewing angle $\theta_v \neq 0$, we
have shown that there should be two breaks in the light curve, the
first one corresponds to the time at which $\gamma\sim
(\theta_j-\theta_v)^{-1}$, and the second one corresponds to the time
when $\gamma\sim (\theta_j+\theta_v)^{-1}$, although these transitions
are very smoothly.

If the jet is not uniform, within which the energy distribution is
given by eq.(10), then the calculation is more complicated. In the case
of $\theta_v =0$, we can give an analytical result, for
$k<8/(p+4)$(where $p$ is the spectral index of electron energy
distribution) there should be two breaks in the light curve correspond
to $\gamma\sim\theta_c^{-1}$ and $\gamma\sim\theta_j^{-1}$
respectively, while for $k>8/(p+4)$ there should be only one break
corresponds to $\gamma\sim\theta_c^{-1}$, after that the flux decays as
$F_\nu\propto T^{-3p/4}$. We argue that this may explain some rapidly
fading afterglows whose light curves have no breaks, since the time
$T_1$, at which $\gamma\sim\theta_c^{-1}$, is usually earlier than our
first observation time.

If the jet is non-uniform and the viewing angle $\theta_v\neq 0$, only
numerical results can be given. We have shown that in this case the
shape of afterglow light curve is dependent on the values of
$\theta_v$, $\theta_c$ and $k$. When $\theta_v$ is near $\theta_c$ and
$k$ is small, then the light curve is similar to the case of
$\theta_v=0$, except the flux is somewhat lower. However, if the
viewing angle $\theta_v$ is larger than $\theta_c$ and $k$ is not
small, then there will be a prominent flattening in the afterglow light
curve, which is quite different from the case of uniform jet, and after
the flattening a very sharp break will be occurred around the time
$\gamma\sim (\theta_v + \theta_c)^{-1}$. We think this is a main
difference between the uniform and non-uniform jet, and we can identify
whether the jet is uniform or not by this feature.

It is not very difficult to understand why sometimes there is a
flattening in the light curve. It is well known that, for a
relativistic blast wave with Lorentz factor $\Gamma\gg 1$, the observer
can only observe a solid angle around $\theta_v$ with a half opening
angle of order $\Gamma ^{-1}$, the contribution from other components
can be neglected, so when the blast wave decelerates, the observer can
see larger components. For the case of non-uniform jet and $\theta_v
>\theta_c$, the energy at $\theta=\theta_v$ is much smaller than that
of smaller $\theta$, so when $\Gamma$ decreases, the observer can
observe more region with larger energy (smaller $\theta$), so there
will be a flattening in the light curve.

We would point out that, in our calculation we have neglected the
sideways expansion of the jet since this process is too complicated.
However, we know that, in fact this process is an important issue in
determining the shape of a light curve, since it will likely
significantly change the shape of the light curve for both the uniform
and non-uniform jets. So we suggest that for a more realistic
calculation the sideways expansion should be taken into account.

\acknowledgements  We are very grateful to J.D. Salmonson for several
important comments that improved this paper. This work is supported by
the National Natural Science Foundation (grants 10073022 and 10225314)
and the National 973 Project on Fundamental Researches of China (NKBRSF
G19990754).

\end{document}